\documentclass[conference]{IEEEtran}
\IEEEoverridecommandlockouts

\usepackage{cite}
\usepackage{amsmath,amssymb,amsfonts}
\usepackage{algorithmic}
\usepackage{graphicx}
\usepackage{textcomp}
\usepackage{xcolor}
\def\BibTeX{{\rm B\kern-.05em{\sc i\kern-.025em b}\kern-.08em
    T\kern-.1667em\lower.7ex\hbox{E}\kern-.125emX}}
\begin{document}

\title{
Validation of an LLM-based Multi-Agent Framework for Protein Engineering in Dry Lab and Wet Lab
}

\author{
\IEEEauthorblockN{
Zan Chen\textsuperscript{1}, 
Yungeng Liu\textsuperscript{1,2}, 
Yu Guang Wang\textsuperscript{1,3}, 
Yiqing Shen\textsuperscript{1,4,*} 
}
\IEEEauthorblockA{
\textsuperscript{1}\textit{Toursun Synbio}, Shanghai, China\\
\textsuperscript{2}\textit{City University of Hong Kong}, Hong Kong, China\\
\textsuperscript{3}\textit{Shanghai Jiao Tong University}, Shanghai, China\\
\textsuperscript{4}\textit{Johns Hopkins University}, Baltimore, USA\\
{\textsuperscript{*}Corresponding authors.} {\footnotesize (Email: \textit{yiqingshen1@gmail.com})}}
}

\maketitle

\begin{abstract}
Recent advancements in Large Language Models (LLMs) have enhanced efficiency across various domains, including protein engineering, where they offer promising opportunities for dry lab and wet lab experiment workflow automation.
Previous work, namely TourSynbio-Agent, integrates a protein-specialized multimodal LLM (\textit{i}.\textit{e}. TourSynbio-7B) with domain-specific deep learning (DL) models to streamline both computational and experimental protein engineering tasks.
While initial validation demonstrated TourSynbio-7B's fundamental protein property understanding, the practical effectiveness of the complete TourSynbio-Agent framework in real-world applications remained unexplored.
This study presents a comprehensive validation of TourSynbio-Agent through five diverse case studies spanning both computational (dry lab) and experimental (wet lab) protein engineering.
In three computational case studies, we evaluate the TourSynbio-Agent's capabilities in mutation prediction, protein folding, and protein design.
Additionally, two wet-lab validations demonstrate TourSynbio-Agent's practical utility: engineering P450 proteins with up to 70\% improved selectivity for steroid 19-hydroxylation, and developing reductases with $3.7\times$ enhanced catalytic efficiency for alcohol conversion.
Our findings from the five case studies establish that TourSynbio-Agent can effectively automate complex protein engineering workflows through an intuitive conversational interface, potentially accelerating scientific discovery in protein engineering.
\end{abstract}


\begin{figure*}[!htbp]
\centerline{\includegraphics[width=0.95\linewidth]{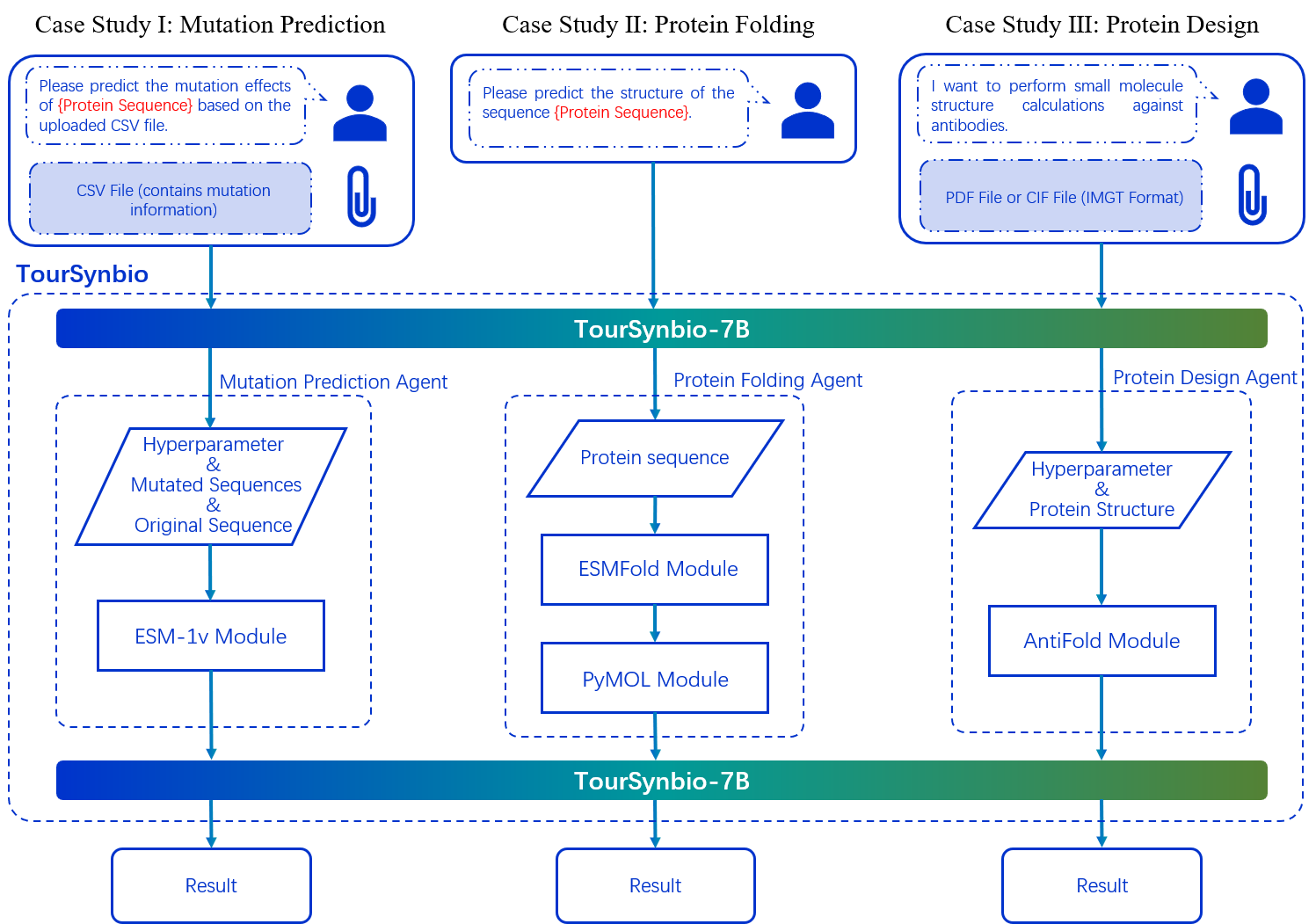}}
\caption{Overview of the TourSynbio-Agent framework for automating protein engineering tasks.}
\label{fig1}
\end{figure*}

\section{Introduction}
Deep learning (DL) has improved the performance and efficiency in protein engineering \cite{gao2020deep, biswas2021low}, such as AlphaFold \cite{jumper2021highly, evans2021protein} and RoseTTAFold \cite{baek2021accurate, humphreys2021computed} achieving much progress in protein structure prediction.
However, the widespread adoption of these DL models to real-world protein engineering workflow remains limited due to their technical complexity, requiring substantial expertise in both protein science and DL for effective implementation \cite{lin2023evolutionary}.
Large Language Models (LLMs) have emerged as promising solutions for interpreting protein-related information, with specialized models like Prot2Text \cite{abdine2024prot2text} and ProteinBERT \cite{brandes2022proteinbert} demonstrating capabilities in processing protein sequences and structures. 
Yet, these protein-specific LLMs have primarily served analytical functions, lacking the ability to autonomously execute complete protein engineering workflows.

To address this limitation, TourSynbio-Agent was recently introduced \cite{shen2024toursynbio} (Fig.~\ref{fig1}), featuring an innovative multi-agent architecture that combines TourSynbio-7B, a protein-specialized multimodal LLM, with domain-specific DL models.
TourSynbio-7B's distinctive capability lies in processing protein sequences directly as natural language, eliminating the need for complex external protein encoders. 
This streamlined approach, coupled with the TourSynbio-Agent's multi-agent design, enables the automated execution of diverse protein engineering tasks through specialized agents.

While initial benchmarking through ProteinLMBench \cite{shen2024fine} demonstrated TourSynbio-7B's fundamental capabilities in protein property analysis, the practical utility of the complete TourSynbio-Agent framework in real-world applications remained unexplored.
This study addresses this gap through five comprehensive case studies spanning both computational (dry lab) and experimental (wet lab) validations.
We first present three computational case studies validating TourSynbio-Agent's ability to handle diverse protein engineering tasks through natural language interactions, including mutation prediction, protein folding, and protein design.
To demonstrate real-world applicability, we then conducted two wet-lab case studies: engineering P450 proteins with up to 70\% improved selectivity for steroid 19-hydroxylation, and developing reductases achieving $3.7\times$ enhanced conversion rates for alcohol compounds.

The major contributions of this work are two-fold.
Firstly, we conduct three dry lab case studies to validate TourSynbio-Agent's capabilities across fundamental protein engineering tasks.
Secondly, we demonstrate the TourSynbio-Agent's effectiveness through two wet-lab-validated case studies.
Together, these results represent the first systematic validation of an LLM-based agent system in real-world protein engineering applications.

\section{Dry Lab Case Study Design}
To validate TourSynbio-Agent's capabilities across fundamental protein engineering domains, we designed three dry lab case studies demonstrating the framework's ability to automate complex workflows through natural language interactions.

\subsection{Case Study \uppercase\expandafter{\romannumeral1}: Mutation Effect Prediction}
Mutation effect prediction is an important component of protein engineering that assesses the functional impact of amino acid substitutions \cite{hopf2017mutation}. 
This computational approach guides rational protein design across multiple applications, including therapeutic development, enzyme engineering, and the analysis of disease-associated mutations.
TourSynbio-Agent streamlines this process by accepting natural language queries (\textit{e}.\textit{g}., ``\textit{predict the effects of mutations in this protein sequence}'') along with protein sequence data in \texttt{CSV} format.
Upon receiving these inputs, TourSynbio-7B activates a specialized mutation prediction agent that leverages the ESM-1v model \cite{meier2021language}. 
The TourSynbio-Agent generates comprehensive outputs including quantitative activity score predictions and qualitative interpretations of mutation effects, enabling researchers to efficiently identify and prioritize promising protein variants.

\begin{figure*}[!t]
\centerline{\includegraphics[width=0.95\linewidth]{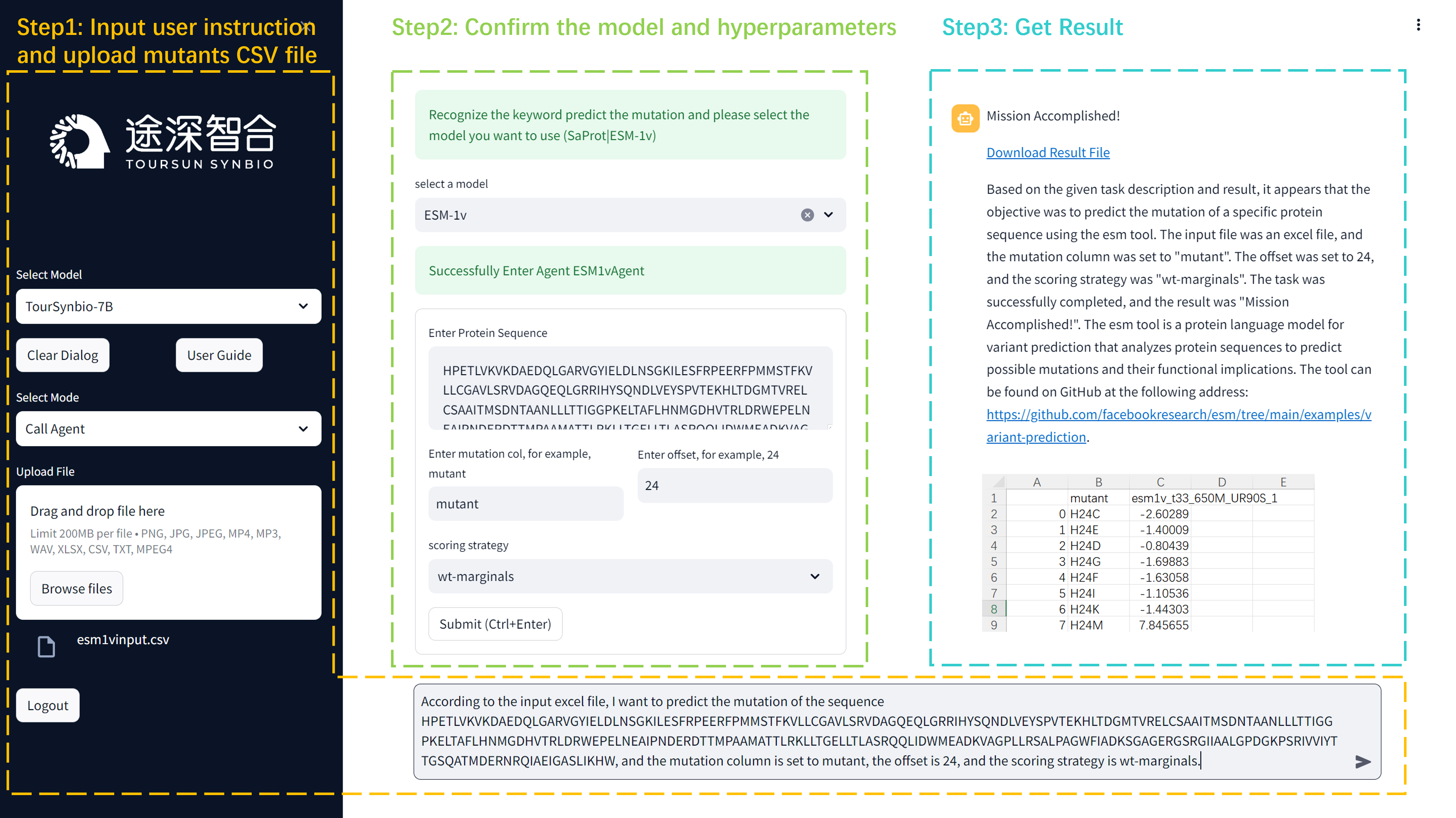}}
\caption{
Workflow of the TourSynbio-Agent mutation prediction pipeline. 
The process consists of three main stages: (1) Input specification, where users provide the protein sequence and upload a \texttt{CSV} file containing mutation information; 
(2) Model configuration, where ESM-1v is selected and parameters including mutation column offset and scoring strategy are defined; and (3) Results generation, displaying predicted activity scores for each mutant in a downloadable format. 
The interface shows the successful prediction of multiple H24 variants, with H24M demonstrating the highest predicted activity score.
}
\label{case1}
\end{figure*}

\subsection{Case Study \uppercase\expandafter{\romannumeral2}: Protein Folding}
The second case study evaluates TourSynbio-Agent's capability to predict three-dimensional protein structures from amino acid sequences.
Users initiate the workflow by submitting a protein sequence alongside a natural language query (\textit{e}.\textit{g}., ``Please predict the structure of the sequence'').
TourSynbio-7B processes this input and activates the protein folding agent, which employs ESMfold \cite{lin2022language} to generate structural predictions.
The predicted structures are visualized through PyMOL \cite{delano2002pymol} and presented to users via an interactive chat interface, with downloadable structure files available for further analysis.

\subsection{Case Study \uppercase\expandafter{\romannumeral3}: Protein Design}
The third case study explores protein design, a more sophisticated task requiring concurrent optimization of structural features and model parameters. 
Users provide design specifications (such as antibody-small molecule interactions) along with structural templates in \texttt{PDB} or \texttt{CIF} format following \texttt{IMGT} standards \cite{lefranc2009imgt}. 
TourSynbio-7B processes these inputs and delegates the task to a specialized protein design agent.
This agent orchestrates a two-step process: first optimizing hyperparameters and processing structural inputs, then utilizing the AntiFold \cite{hoie2023antifold} module to generate designs that meet specified constraints. 
The framework returns complete protein designs optimized for the intended application, whether experimental validation or therapeutic development.

\begin{figure*}[!t]
\centerline{\includegraphics[width=0.95\linewidth]{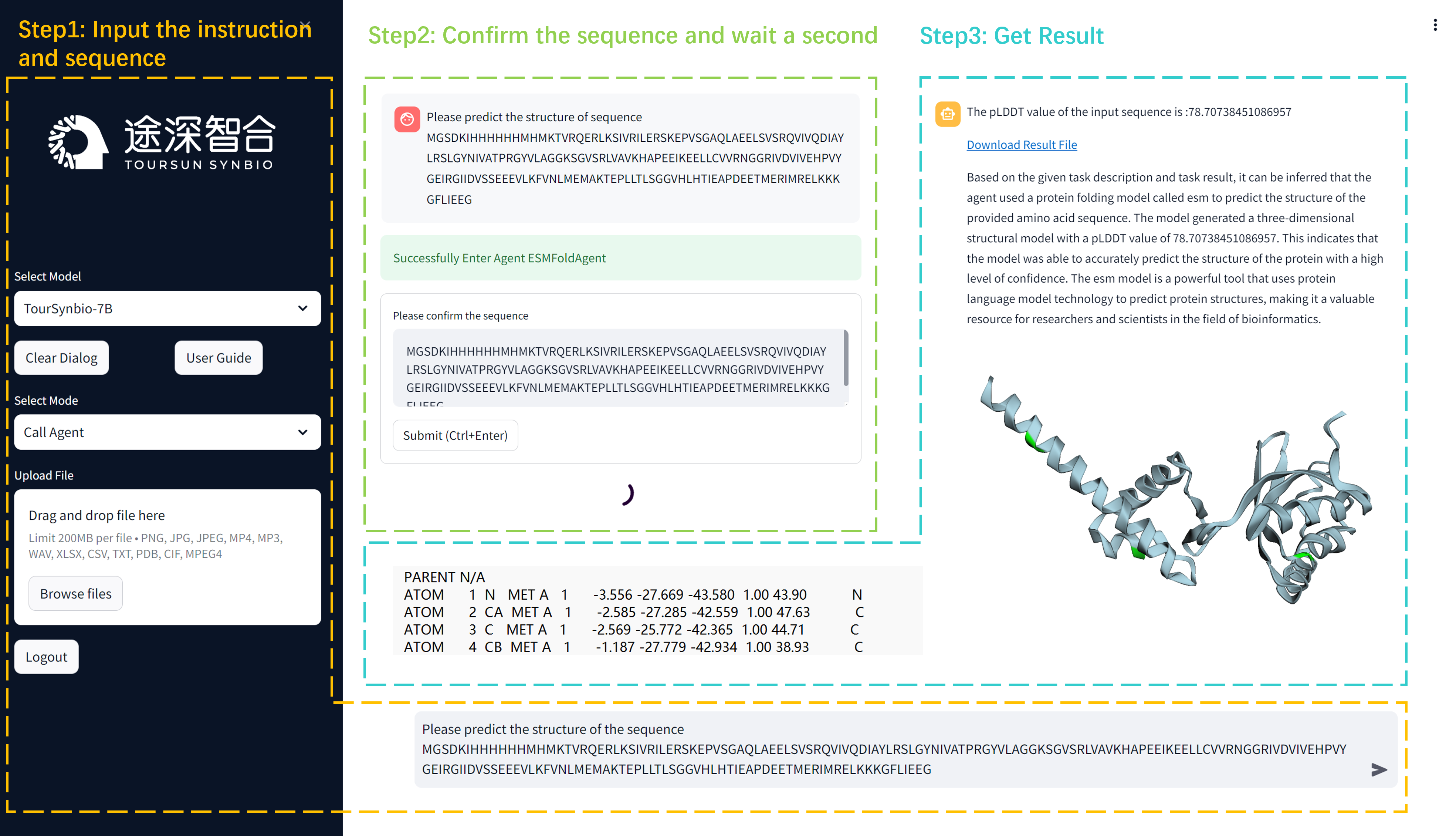}}
\caption{Workflow of the TourSynbio-Agent's protein structure prediction pipeline. 
The process comprises three key stages: (1) Initial setup, where users input the protein sequence and select the ESMfold prediction agent; 
(2) Sequence confirmation and model execution, showing the input protein sequence and the ESMfold processing interface; 
and (3) Results visualization, displaying both the predicted 3D structure in cartoon representation and atomic coordinates in \texttt{PDB} format. 
The interface reports a pLDDT confidence score of 78.7073\%, 
indicating high prediction reliability. The predicted structure shows a mixed $\alpha/\beta$ fold topology with well-defined secondary structure elements, and the coordinate section demonstrates the detailed atomic-level output generated by the model.}
\label{case2}
\end{figure*}

\section{Dry Lab Case Study Results}
We evaluated TourSynbio-Agent's performance across three fundamental protein engineering tasks: mutation prediction, protein folding, and protein design. 
For each case study, we present detailed analyses of the framework's capabilities, including input processing, computational predictions, and result interpretation.

\subsection{Experimental Process \uppercase\expandafter{\romannumeral1}: Mutation Prediction}
We validated TourSynbio-Agent's ability to predict mutation effects on protein activity, an important capability for protein engineering applications such as therapeutic development and enzyme optimization. 
The study workflow, illustrated in Fig.~\ref{case1}, consists of three distinct stages: input specification, model configuration, and results analysis.
The prediction pipeline requires two primary inputs: (1) the wild-type protein sequence, which in this case study was "\textit{HPETLVKVKDAEDQLGARVGYIELDLNSGKILESFRPEERFMMSTFKV...}" and (2) a structured dataset in \texttt{CSV} format containing a library of single and/or multiple point mutations derived from the original sequence. 
The mutation information was specified in the ``mutant'' column of the input file, with an offset parameter of 24 to correctly align mutation positions with the protein sequence.
Using the TourSynbio-7B interface, we configured the mutation prediction agent to utilize ESM-1v, a protein language model specifically trained for mutation effect prediction. 
The scoring strategy was set to ``wt-marginals'' to compute the relative impact of mutations compared to the wild-type sequence. 
This configuration enables the ESM-1v to analyze how each mutation affects protein stability and function relative to the original sequence.
The ESM-1v then evaluated each variant in the mutation library, generating activity scores that quantify the predicted functional impact.
Our analysis focused on mutations at position H24, examining multiple variants including H24E (-1.40009), H24D (-0.80439), H24G (-1.69883), and H24M (7.84565). 
Among these variants, H24M exhibited the highest activity score of 7.84565, suggesting an enhancement in protein performance compared to the wild-type sequence.
The output, provided in a downloadable format, includes detailed scores for each mutation variant, enabling researchers to prioritize promising mutations for experimental validation. 
This computational screening approach demonstrates how LLMs can accelerate the protein engineering cycle by identifying high-potential variants before laboratory testing.

\subsection{Experimental Process \uppercase\expandafter{\romannumeral2}: Protein Folding}
This study evaluated TourSynbio-Agent's capability to predict protein three-dimensional structures through an automated pipeline. 
The experimental workflow, illustrated in Fig.~\ref{case2}, demonstrates the seamless integration of state-of-the-art structure prediction methods into an accessible framework.
The prediction pipeline begins with sequence input through TourSynbio-Agent's conversational interface. 
In this case study, we analyzed a protein sequence starting with "\textit{MGSDKIHHHHHHHMHKMTVRQERLKSIVRILER...}", which was directly input through the conversational interface. 
Upon sequence submission, TourSynbio-7B activated its ESMfold Agent, a specialized model that performs end-to-end atomic-level structure prediction without requiring multiple sequence alignments or template structures.
The ESMfold generates both atomic coordinates and confidence metrics. 
For our test sequence, the ESMfold achieved a pLDDT (Predicted Local Distance Difference Test) score of 78.7073\%, indicating substantial confidence in the predicted structure's accuracy. 
%
The output is presented in two complementary formats: (1) a detailed \texttt{PDB} file containing atomic coordinates (exemplified in the figure by entries such as "\textit{ATOM 1 N MET A 1 -3.556 -27.669 -43.580 1.00 43.90 N}"), and (2) an interactive visualization interface showing the predicted structure in cartoon representation. 
The structural model reveals a mixed $\alpha/\beta$ fold topology with well-defined secondary structure elements, allowing immediate visual assessment of key structural features.
This automated structure prediction pipeline streamlines what has traditionally been a complex and computationally intensive process. 
The combination of high-confidence predictions, detailed atomic coordinates, and instant visualization capabilities demonstrates TourSynbio-Agent's potential to accelerate structure-based research workflows in both academic and industrial settings.

\begin{figure*}[!t]
\centerline{\includegraphics[width=0.95\linewidth]{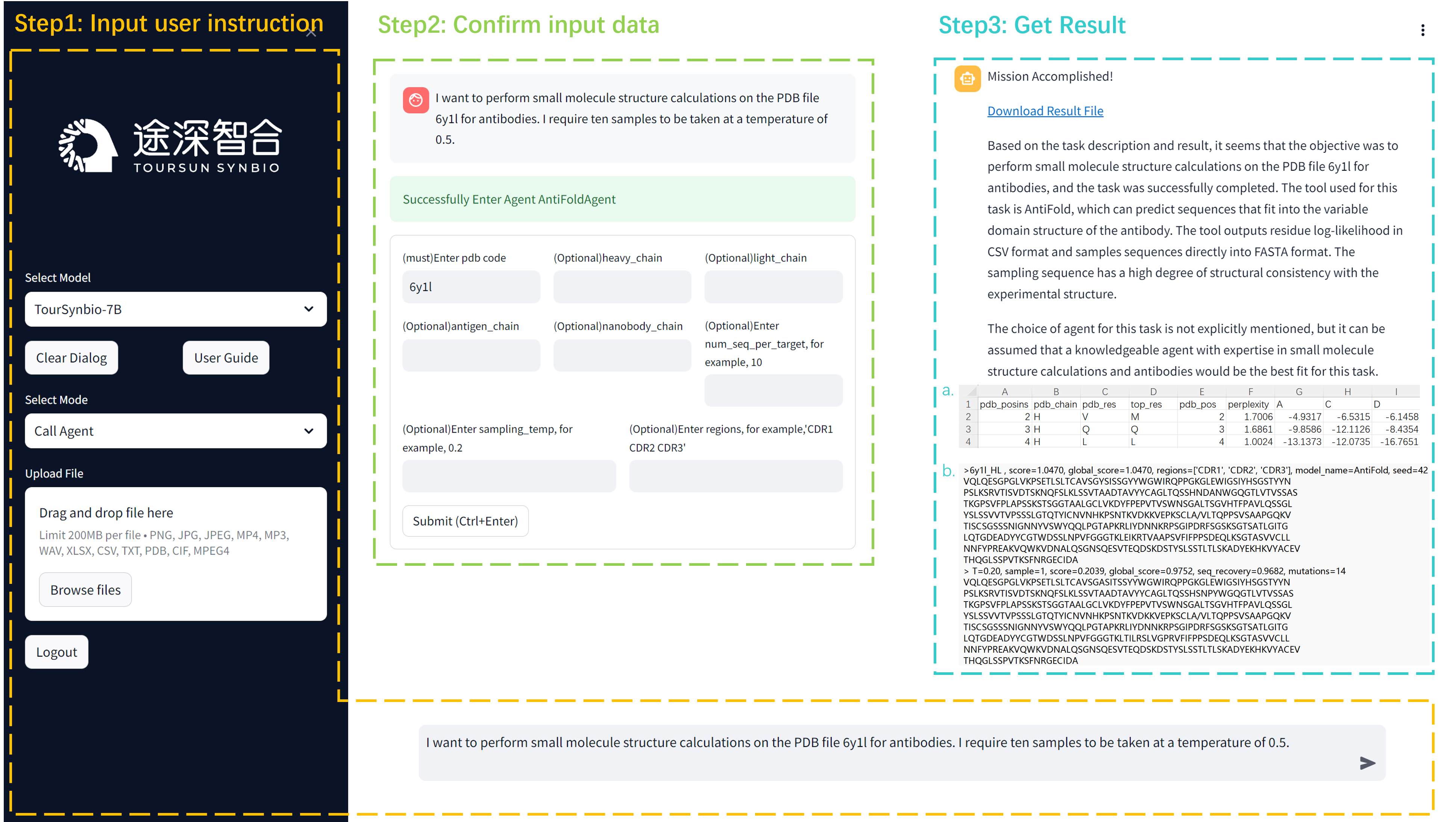}}
\caption{Workflow of TourSynbio-Agent's antibody design pipeline using Antifold. 
The process consists of three stages: (1) Initial configuration, where users select the model and specify the \texttt{PDB} input (6y1l); 
(2) Parameter specification, showing input fields for structural components (heavy chain, light chain, antigen chain, nanobody chain) and sampling parameters (temperature of 0.5, CDR regions); 
and (3) Results presentation, displaying both a tabular output with structural scores and generated antibody sequences in \texttt{FASTA} format. The conversational interface allows for precise control over the sampling process while maintaining ease of use. 
The output panel shows multiple sampled sequences with their associated scores, demonstrating the model's ability to generate structurally consistent antibody variants.}
\label{case3}
\end{figure*}

\subsection{Experimental Process \uppercase\expandafter{\romannumeral3}: Protein Design}
This experiment leveraged TourSynbio-Agent's capabilities to explore sequence variations in an antibody structure (PDB ID: 6y1l) while maintaining its structural integrity and functional properties. 
The workflow, illustrated in Fig.~\ref{case3}, demonstrates the integration of Antifold's inverse folding capabilities into a systematic antibody design pipeline.
The study process began with the specification of input parameters through TourSynbio-Agent's conversational interface. 
Users input the PDB code ``6y1l'' and can optionally specify structural components including heavy chain, light chain, antigen chain, and nanobody chain identifiers. 
The sampling parameters were configured with a temperature of 0.5 and specific complementarity-determining regions (CDRs) targeted for design. 
Upon parameter confirmation, TourSynbio-7B activated the Antifold Agent to perform structure-based sequence calculations.
The framework generated comprehensive results in two complementary formats. The first output, provided in \texttt{CSV} format, delivered a detailed residue-level analysis containing position-specific data including chain identifiers (\textit{e}.\textit{g}., H, L), original and predicted residue identities, and structural metrics such as per-residue perplexity.
The analysis revealed varying structural compatibility scores across different positions, with chain H positions showing scores ranging from -4.9317 to -16.7651, providing quantitative insights into the structural impact of mutations.
The second output format presented sequence sampling results in \texttt{FASTA} format, preserving the original antibody sequence as a reference while generating multiple design variants. 
Each variant was accompanied by detailed scoring metrics, with the exemplar design achieving a global score of $1.0470$, indicating strong structural consistency. 
The framework evaluated specific CDR regions and provided additional metrics including a sequence recovery rate of 0.9682 and a mutation count of 14, enabling researchers to assess both local and global impacts of the designed variations.

This study demonstrates TourSynbio-Agent's capability to handle sophisticated protein engineering tasks. The framework efficiently generates both detailed residue-level predictions and complete sequence variants, providing researchers with quantitative metrics to evaluate structural stability and functional potential.

\section{Wet-lab Case Study}
To validate TourSynbio-Agent's practical utility in real-world applications, we conducted two experimental case studies focusing on enzyme engineering. 
These studies demonstrate the framework's ability to optimize enzyme properties through iterative computational prediction and experimental validation cycles.

\subsection{Wet-lab Study \uppercase\expandafter{\romannumeral1}: Enhancing Steroid Compound Selectivity}

Steroid compounds represent a crucial class of bioactive molecules that serve essential physiological functions, from maintaining cellular membrane integrity to acting as hormonal signaling molecules \cite{carson1990steroid}. 
Their therapeutic applications span multiple medical domains, including cardiovascular \cite{demer2018steroid} and cerebrovascular diseases \cite{witt2014steroids}. 
This case study focused on engineering cytochrome P450 enzymes to enhance their selectivity for steroid 19-hydroxylation.
While P450-catalyzed reactions typically generate multiple products, only one specific hydroxylation product possesses the desired therapeutic properties. 
Our objective was to achieve a 70\% improvement in selective product formation while maintaining catalytic activity, which is a threshold requirement for industrial-scale implementation.

\begin{figure}[h!]
    \centering
    \centerline{\includegraphics[width=\linewidth]{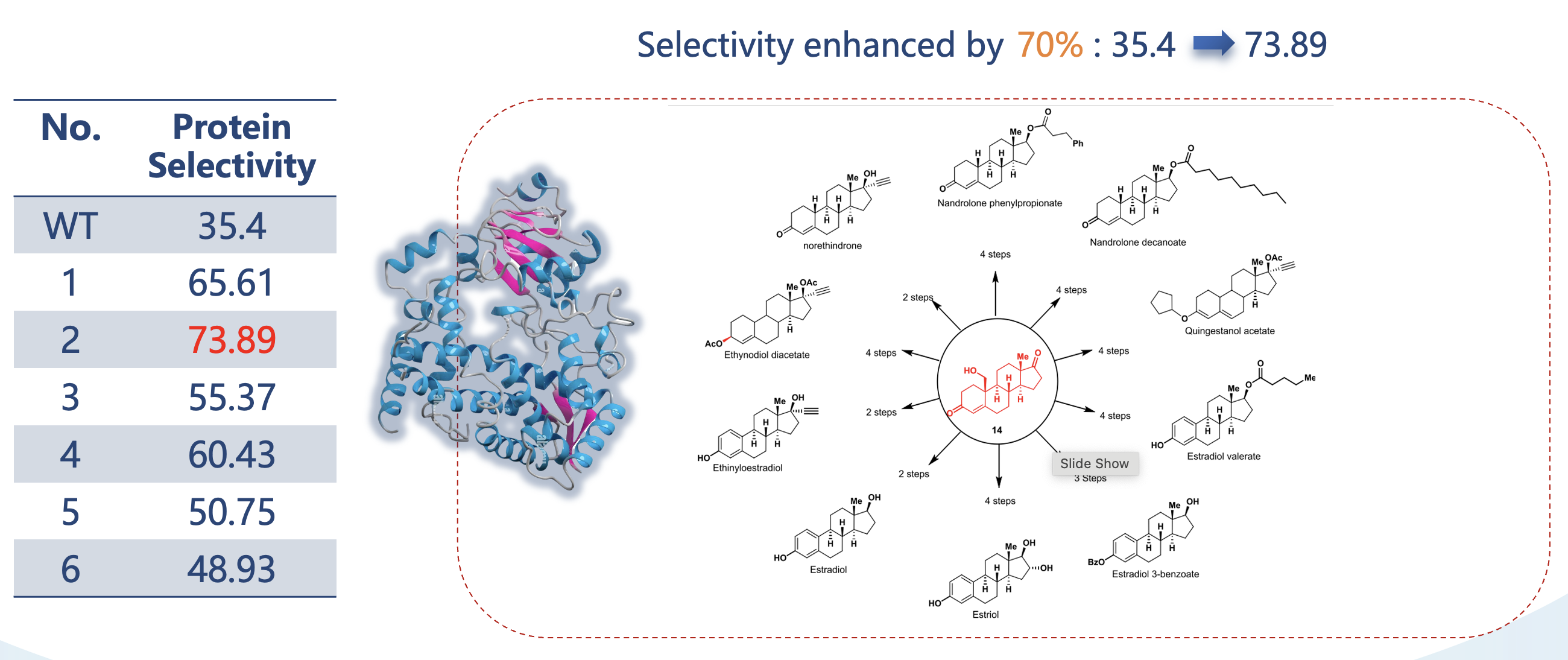}}
    \caption{The goal is to modify the P450 protein, which catalyzes the 19-hydroxylation of steroid compounds, to increase its selectivity by 70\% for the effective product, a crucial step for scaling up production efficiency.}
    \label{fig:case2}
\end{figure}

\subsubsection{Engineering Strategy and Implementation}
The engineering process proceeded in two distinct phases. 
During the initial screening phase, TourSynbio-Agent generated 200 single-site mutation candidates within two weeks, followed by a three-week experimental validation period to collect comprehensive activity and selectivity data. 
In the subsequent focused optimization phase, this experimental data was used to fine-tune the prediction models. 
TourSynbio-Agent then generated 10 optimized variants containing up to five mutations each, which underwent detailed experimental characterization for both selectivity and activity.

\subsubsection{Results}
The engineering campaign yielded outcomes that validated TourSynbio-Agent's effectiveness in protein engineering, as shown in Fig.~\ref{fig:case2}. 
The framework demonstrated strong predictive accuracy, achieving a correlation coefficient of 0.7 between computational predictions and experimental measurements. 
Most notably, the best-performing variant achieved the target 70\% improvement in product selectivity while maintaining robust catalytic activity. 
These performance metrics met the stringent criteria for potential industrial implementation, highlighting TourSynbio-Agent's capability to address complex biocatalysis optimization challenges.

\subsection{Wet-lab Study \uppercase\expandafter{\romannumeral2}: Assisting customers in enhancing the catalytic conversion rate of enzymes}

\begin{figure}[h!]
    \centering
    \centerline{\includegraphics[width=\linewidth]{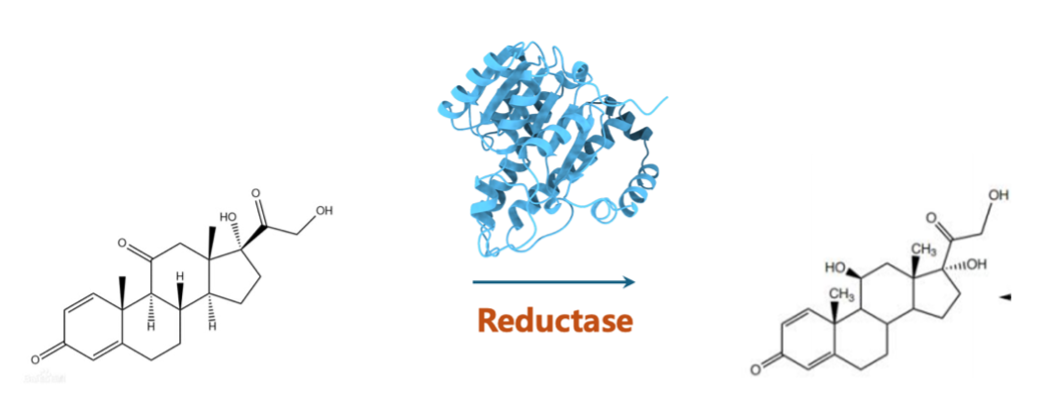}}
    \caption{Reductase catalysis of alcohol compounds.}
    \label{fig:case3}
\end{figure}

Steroid hormones and their synthetic derivatives represent an important segment of the pharmaceutical industry, with applications spanning reproductive health, metabolic disorders, inflammatory conditions, and immunological diseases \cite{holsboer2010stress}. 
Key compounds in this category include progesterone, testosterone, estradiol, cortisol, and aldosterone, along with various synthetic progestogens. 
The growing prevalence of age-related and lifestyle diseases has driven increasing demand for these therapeutic agents, necessitating more efficient production methods \cite{fehlings2015aging, christiansen2016growth}.
This case study focused on optimizing reductase enzymes to enhance their catalytic efficiency in alcohol compound synthesis. 
Improving catalytic conversion rates directly impacts manufacturing productivity and economic viability by maximizing product formation within fixed reaction timeframes. 
The engineering objective was to increase the enzyme's catalytic efficiency while maintaining product specificity as shown in Fig.~\ref{fig:case3}.

\subsubsection{Engineering Approach}
The optimization process began with a dataset comprising the wild-type reductase sequence and activity measurements for 29 single-point variants. 
TourSynbio-Agent analyzed this initial dataset to develop structure-function relationships and subsequently recommended 10 novel single-point mutations predicted to enhance catalytic performance. 
These engineered variants underwent comprehensive experimental validation over a four-week period.

\subsubsection{Results}
The reductase engineering campaign demonstrated both the predictive accuracy of TourSynbio-Agent and its ability to achieve substantial functional improvements. 
The framework's predictions showed a strong correlation with experimental results, achieving a correlation coefficient of 0.7 between computational predictions and measured activities. 
This validation confirms TourSynbio-Agent's reliability in identifying beneficial mutations for enzyme optimization.
Among the designed variants, the most successful candidate exhibited a $3.7\times$ enhancement in catalytic conversion rate compared to the wild-type enzyme. 
This improvement in catalytic efficiency translates directly to practical benefits: increased product yields, reduced reaction times, and more efficient utilization of raw materials. 
%
%

\section{Conclusion and Discussion}
This study presents a comprehensive validation of TourSynbio-Agent through five diverse case studies, demonstrating its effectiveness in automating complex protein engineering workflows. 
The three computational case studies showcase the TourSynbio-Agent's ability to streamline traditionally complex tasks through an intuitive natural language interface. 
The successful wet-lab validations, particularly the engineering of P450 proteins with 70\% improved selectivity and reductases with 3.7× enhanced catalytic efficiency, provide concrete evidence of TourSynbio-Agent's practical utility in real-world applications.
The integration of a protein-specialized multimodal LLM with domain-specific agents enables TourSynbio-Agent to bridge the gap between computational predictions and experimental implementation. 
By providing researchers with actionable insights and automated workflow management, the framework reduces the technical barriers typically associated with advanced protein engineering techniques.

Several directions emerge for future development. 
First, establishing standardized evaluation metrics specifically designed for LLM-based protein engineering frameworks would enable systematic comparison of different approaches and facilitate continued improvement. 
These metrics should assess both computational accuracy and practical utility in experimental settings.
Second, expanding the TourSynbio-Agent's knowledge base and integrated datasets would enhance its capabilities across a broader range of protein engineering applications, from therapeutic antibody design to industrial enzyme optimization.
Finally, investigating the framework's potential for autonomous experimental design and optimization could further accelerate the protein engineering cycle.

\bibliographystyle{IEEEtran}
\bibliography{citations.bib}

\end{document}